**This is the Author Accepted Manuscript (AAM) of:**

Nicole Holzmann, *From Promise to Practice: Closing the Application Gap in Quantum Computing*,

Philosophical Transactions of the Royal Society A, accepted for publication, 2026.

# From Promise to Practice: Closing the Application Gap in Quantum Computing

**Nicole Holzmann*†**

*No affiliation*

## Abstract

Quantum computing is a deep technology whose progress cannot be driven effectively from one direction alone. While the field has developed a growing catalogue of mathematically grounded algorithmic speedups, industrial impact will depend just as much on starting from real industrial decision contexts and working downward to what must be computed, validated and integrated. In this Perspective, I argue that sustained progress requires treating these two directions: bottom-up development from physics, hardware and algorithms, and top-down development from industrial needs and constraints. Equally primary and continuously coupled.
This dual-viewpoint is not a matter of balance for its own sake. Quantum computers cannot solve arbitrary problems, so engagement with industry must remain anchored in algorithmic tractability. Yet tractable computations are rarely valuable unless they connect to decision points in established workflows such as candidate selection in drug discovery or the design of a new aircraft shape with improved aerodynamics. I analyse how historical narratives and structural separations of expertise slowed the formation of this coupling and outline what it takes to build it: explicit interfaces between technical teams and domain context and intermediate layers that translate quantum outputs into decision-relevant observables without suffocating foundational innovation. Framed this way, quantum computing's opportunity is clearest where deep physical modelling meets high-value decisions. Provided the field co-designs both sides from the outset.

## Introduction

The current discourse around quantum computing is often characterised by a tension between extraordinary promise and persistent scepticism. On the one hand, a growing number of quantum algorithms have been developed that offer theoretical advantages for specific classes of problems. On the other hand, translating these advances into practical, industrially relevant applications remains a central challenge. This tension is not unique to quantum computing. In fact, it closely mirrors the early history of classical computing.

It is often noted and usually not without a degree of irony that early commentators on computing in the mid-20th century made famously pessimistic predictions about its eventual impact. Statements such as the claim that there would be a market for only "a handful" of computers worldwide are frequently cited as emblematic examples of technological short-sightedness. While many of these quotations are either apocryphal, taken out of context or were never meant in the way they are commonly repeated, they nonetheless convey something important about the prevailing sentiment of the period. Early computers were rare, expensive, unreliable and extraordinarily difficult to program.

The ENIAC (Electronic Numerical Integrator and Computer) [1] devised in the 1940s occupied a surface area of roughly 170 $m^2$ and weighed approximately 30 tons. It relied on about 18,000 vacuum tubes and failures were sufficiently frequent. Even a single malfunction could halt a computation or invalidate results. ENIAC was not a stored-program machine, its operations were fundamentally limited to numerical arithmetic: addition, subtraction, multiplication, division and square roots, manually configured to perform specific tasks such as ballistic trajectory calculations. From this vantage point, it was genuinely difficult to imagine a world in which computation would become ubiquitous, embedded in everyday devices and indispensable across nearly every industrial sector.

*Author for correspondence (nicole.holzmann@googlemail.com)

†Present address: Guentherstrasse 40, 60528 Frankfurt am Main, Germany.

This historical perspective offers a useful corrective to today's debates about quantum computing. The fact that current quantum devices appear limited, fragile and poorly matched to real-world needs does not in itself imply that they will remain so. However, the success of classical computing was not inevitable, nor was it driven by abstract theoretical insight alone. It was shaped by sustained investment, evolving use cases and, critically, the gradual alignment between what machines could do and what society and industry needed them to do.

It is not sufficient to catalogue what early classical computers were capable of. It is more instructive to examine why their early capabilities proved such a poor guide to their eventual impact.

This lesson extends beyond early general-purpose CPUs to later waves of computing hardware, where similar patterns reappeared under different architectural constraints. A particularly instructive example of this dynamic can be found in the evolution of graphics processing units (GPUs) [2]. GPUs were designed around architectural principles fundamentally different from those of central processing units, prioritising high-throughput execution and massive parallelism over low-latency, sequential control. In computational approaches formulated around CPU-centric abstractions, algorithms are expressed as largely sequential operations. As a consequence, they were naturally ill-suited to GPU architectures. Applied without reformulation such approaches tend to underutilise parallel resources, incur substantial data-movement overheads and in many cases result in longer runtimes rather than acceleration.

The eventual impact of GPUs emerged only because algorithms and workflows were rethought to align with these architectural characteristics. This evolution did not occur abruptly, but built on earlier developments in parallel computing and vectorised execution, including SIMD (Single Instruction, Multiple Data) paradigms on CPUs. GPUs first gained prominence in graphics and gaming, where parallelism was inherent to the problem structure. An area where GPUs were subsequently adopted will be familiar to many in the scientific community: molecular dynamics simulations where force calculations over numbers of particles naturally map onto parallel execution, speeding up e.g. protein simulations so significantly that it unlocked larger system sizes and longer timescales. Only later did GPUs become central to artificial intelligence and machine learning, where hardware architectures, software frameworks and algorithmic approaches co-evolved to exploit GPU capabilities.

This push in the artificial intelligence realm coincided with a broader structural shift in where and how application-driven progress occurred. As deep learning matured, impact increasingly depended on the alignment of three key inputs: large-scale computing resources, access to vast datasets and the ability to attract and retain highly specialised technical talent. Industry came to dominate all three [3].

This translated directly into research outcomes, ultimately enabling breakthroughs such as AlphaFold [4]. Industrial actors not only drove the development of the most capable models and software frameworks, but also became increasingly prominent in academic publications, benchmark-setting results and the definition of what constituted state-of-the-art performance [5,6]. In effect, the centre of gravity for progress moved from isolated advances to ecosystems capable of sustaining end-to-end innovation at scale. The leading role of the GPU industry, with NVIDIA at the forefront [7], was not accidental. Beyond strong commercial incentives, these companies drew on experience from their gaming origins, where success depended on shaping hardware architectures that could support demanding, rapidly evolving software workloads and serve a large and exacting user base.

The GPU story illustrates more than a lesson about architectural suitability. It shows that transformative impact arises when hardware capability, algorithmic paradigms, data availability and organisational capacity evolve together and when institutions exist that can integrate these elements coherently.

The lesson from GPUs extends beyond architecture. Their eventual impact was not driven solely by advances in hardware, but also by the emergence of an ecosystem capable of connecting hardware development, software tooling and application needs end to end. Libraries, programming frameworks, communities of developers and application-specific software stacks created pathways through which computational capabilities could be translated into practical outcomes. A similar institutional challenge exists for quantum computing. The question is therefore not only whether quantum computations can be performed, but also whether the organisational structures, expertise, standards and interfaces required to connect quantum-computed quantities to industrial decision-making can emerge alongside them. In this sense, the GPU experience highlights an institutional as well as a technical lesson: transformative impact depends on the co-development of hardware, software and application ecosystems, rather than on hardware capability alone.

In quantum computing, considerable effort has gone into identifying promising directions and problem classes where quantum approaches could offer advantages, yet clear demonstrations of industrial impact have so far not emerged. This invites reflection on what lessons can be learned from the remarkable trajectory of classical computing, which ultimately succeeded in turning technical capability into broad and durable value.

What if quantum computing succeeds scientifically, but not industrially? What if fault-tolerant quantum computers are ultimately realised, yet applications with clear industrial benefit are never identified or developed? In such a scenario, the value of quantum computers and algorithms would not vanish. They would remain of significant academic interest, enabling new insights in quantum chemistry and physics and supporting high-impact scientific research but not sufficiently beyond scientific value to impact industry broadly.

Industrial engagement depends on the prospect of *decision-relevant impact*: the ability of a technology to measurably influence real decisions such as go/no-go choices, design trade-offs, risk assessments or resource allocation. This can be achieved by reducing uncertainty, shortening development timelines, lowering cost or risk or enabling options that were previously inaccessible. Developing credible pathways to such influence on decision-making is essential to move beyond the technology's theoretical potential and to justify sustained private investment. Large-scale scientific infrastructure provides a useful analogy. National facilities, such as neutron sources operated by organisations like the UK's Science and Technology Facilities Council, deliver extraordinary value to the research community through access to highly specialised instruments [8,9]. Their success relies on public funding and a well-defined scientific user base, rather than on widespread industrial adoption or commercial return. From such a starting point, the development of industrial impact remains possible, but it tends to proceed at a markedly slower pace, constrained by limited access, long proposal cycles and the absence of sustained, application-driven investment.

Unlike neutron sources, however, much of today's quantum computing effort is driven by commercial companies. This creates both a requirement and an opportunity: to develop applications that connect quantum capabilities to real workflows and industrial outcomes. Framed in this way, the application gap is not a sign of limitation, but a defining challenge for the next phase of the field. Those organisations that recognise its importance and invest in closing it will be best positioned to lead as quantum computing matures.

Recent work has framed the challenge of quantum applications in terms of a multi-stage pipeline from algorithm discovery to deployment, stressing the need to bridge abstract speedups with real-world use cases and resource estimation while proposing a "problem-first" approach that builds from algorithmic advantage upwards and a roadmap toward quantum advantage for industry-grade applications [10]. In other studies, the implications that quantum computing may have in coming years has been discussed [11,12] or on alleviating the hype [13,14], but while these works have highlighted the need to bridge the gap between quantum algorithm development and practical deployment, they largely remain at the level of creating benefit by finding problems directly fitting to algorithmic advantage. This perspective builds on that discussion but shifts the focus from the existence of quantum advantage to the problem of preserving that advantage as computational outputs are translated into decision-relevant industrial quantities. It treats quantum computing as a deep-technology challenge in which progress must occur simultaneously and iteratively from the side of algorithm development and said translation layer and the side of industrial requirements and benefit.

In this perspective, the aim is to contribute a complementary viewpoint rooted in strategic reflection on where conceptual changes are needed if quantum computing is to yield applications of industrial benefit. Understanding where progress has stalled requires stepping back and asking why. Drawing on insights gained through years of engagement in quantum technology development, I reflect on the current state of the field from a position that is both invested and pragmatic. From an algorithmic standpoint, the potential of quantum computing is undeniable and progress in both hardware and algorithms continues steadily. The lack of clear application pathways reveals a deeper issue in the wider quantum community: A broader disconnect with the intrinsic complexity of translating abstract quantum advantage into real-world utility proven greater than early narratives could readily accommodate. It is from this underestimation that critical gaps have emerged. Gaps that must first be acknowledged before they can be meaningfully addressed.

This perspective is shaped by the author's background in quantum chemistry and by several years spent exploring potential industrial applications of quantum computing. As no individual can engage with all areas of quantum computing with equal technical depth, the discussion draws primarily on examples from chemistry, where the author has the strongest domain expertise. These examples are used not because chemistry is necessarily the only or even the ultimate destination for industrial quantum advantage, but because they provide a concrete and technically grounded setting in which to examine the broader challenge addressed in this article: how quantum computational capabilities can be translated into industrially relevant outcomes.

# From Quantum Computation to industrial Benefit

**(a) Algorithmic Promise**

From the technology perspective, quantum algorithms span a wide range of asymptotic advantages. Problems such as integer factorisation and the simulation of electronic structure in quantum chemistry are often highlighted as prime examples where quantum algorithms offer exponential speedups over the best known classical approaches (Figure 1). In both cases, the advantage arises from a close correspondence between the computational problem and quantum computation itself. For integer factorisation, this correspondence is algorithmic in nature, whereas in quantum chemistry it reflects a direct mapping of interacting many-electron systems onto quantum states, whose classical description scales exponentially with system size.

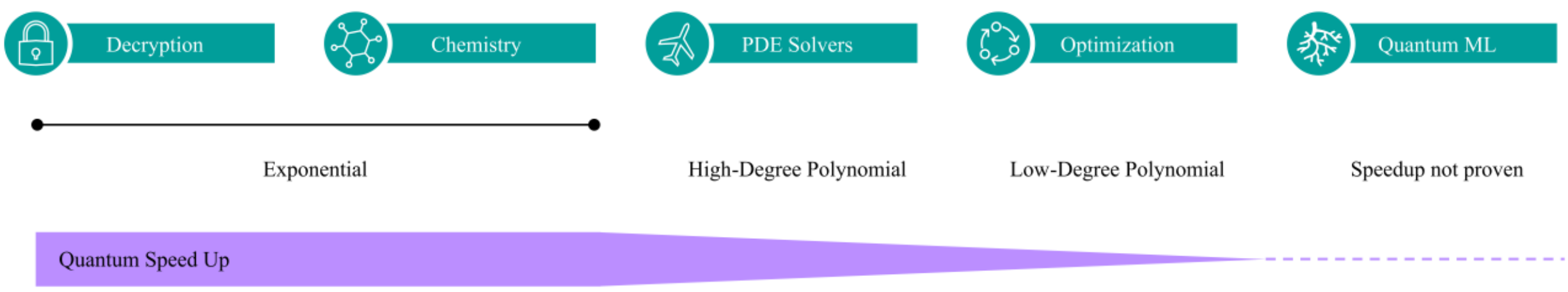

Figure 1: Spectrum of quantum algorithm speedup. Examples shown are representative rather than exhaustive. Speedups depend on the specific problem formulation, algorithm and computational assumptions and do not necessarily translate directly into practical performance.

**Decryption Algorithms**

Integer factorisation has a very straightforward application in decryption of RSA while Shor's discrete log algorithm can be used for or ECC (Elliptic Curve Cryptography) encrypted data. The reason why this has been quickly and undoubtedly recognised as an application is related to the immediate ability of effectively compromising an RSA or ECC key [15]. However, this clean connection between a quantum algorithm and a practical outcome is the exception. The scope for decryption is limited and time-bounded: as cryptographic standards transition to post-quantum schemes [16], the practical utility of large-scale decryption diminishes. Beyond cryptanalysis, integer factorisation does not naturally give rise to a broader class of industrial applications.

**Chemistry Algorithms**

This leads to quantum chemistry often being cited as one of the most promising application areas for quantum computing, owing to the exponential scaling of exact classical electronic-structure methods. In this context, the frequently cited exponential advantage of quantum algorithms is defined relative to highly accurate classical approaches such as Full Configuration Interaction (Full-CI), which provide exact solutions within a chosen basis but become intractable beyond very small systems.

This comparison is scientifically well founded and it is precisely this regime that algorithms such as Quantum Phase Estimation (QPE) are designed to address [17]. QPE provides a systematic and controllable route to eigenvalues of molecular Hamiltonians by separating two distinct aspects of the calculation: how faithfully the quantum system itself is represented and how precisely the corresponding energy is extracted. In practice, this separation is realised through two groups of qubits. One group (often referred to as the *state register)* encodes the electronic wavefunction of the system, determining how accurately the quantum state can be represented. A second group (commonly called *ancilla* qubits) is used to resolve the energy associated with that state, with their number setting the precision of the energy estimate. (Figure 2).

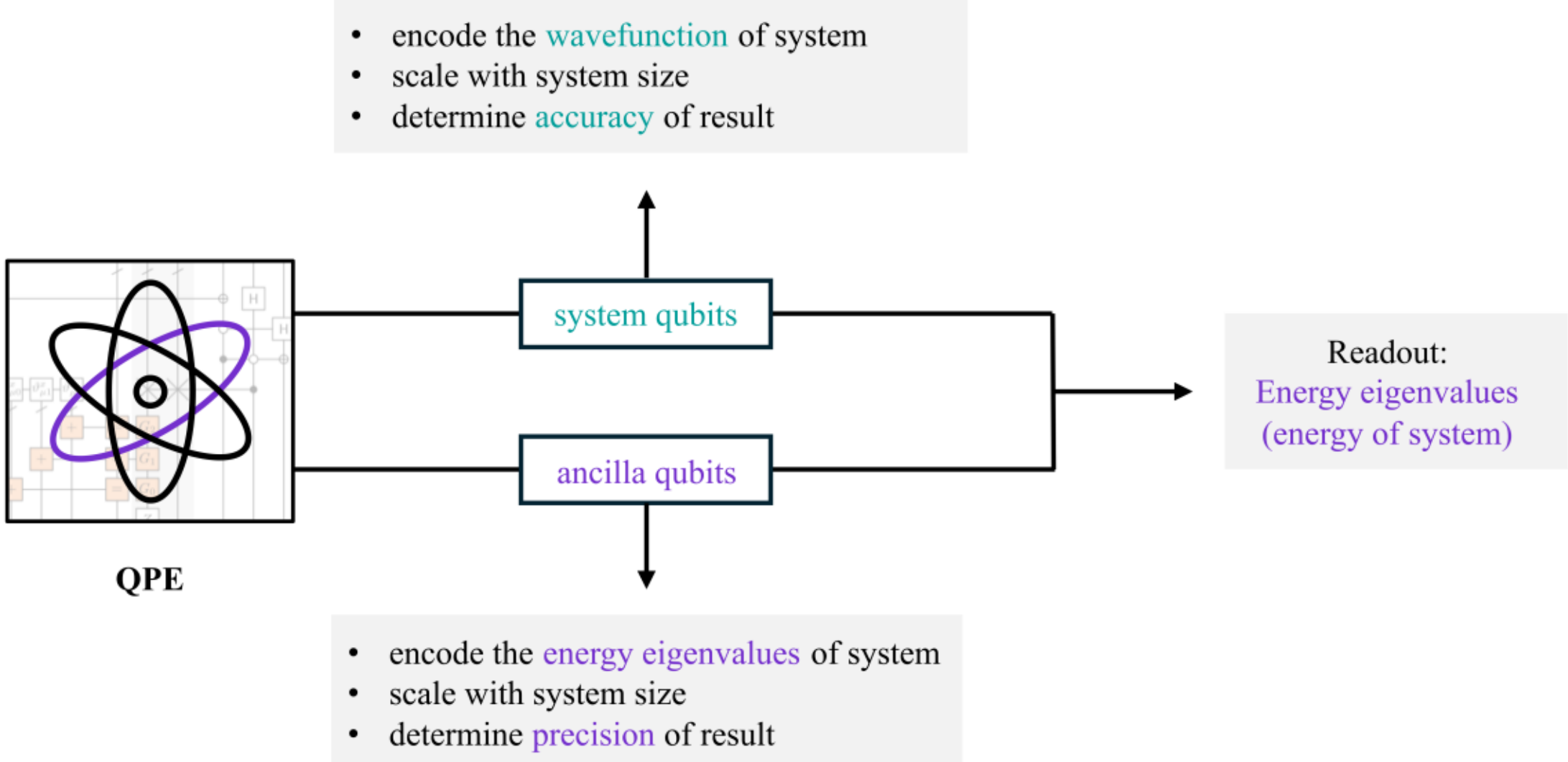

Figure 2: Quantum Phase Estimation algorithm, a simplified view, shown as one representative example.

Both accuracy and precision can thus be controlled, but only at the cost of increased quantum (and associated classical) resources. Within a fault-tolerant framework, these trade-offs can be managed in a predictable and systematic manner. For this reason, QPE represents the most rigorous realisation of exponential quantum advantage in chemistry. At the same time, its reliance on fault-tolerant quantum computing places it firmly on a longer-term timescale.

However, Full Configuration Interaction is not the benchmark against which most industrial chemistry problems are solved today. In practice, approximate but far more efficient classical methods (for electronic calculations most prominently density functional theory) are widely used. While DFT sacrifices formal exactness, it often provides sufficient accuracy at a fraction of the computational cost. FCI cannot be applied using conventional computers at the size of systems that are typically required, as this leads to practically infinite runtimes.

As I will discuss later in this perspective, a substantial number of industrial workflows actually do not rely on electronic-structure calculations at all. Instead they employ empirical models, classical force fields or even relatively simple statistical and regression-based approaches to guide design. These methods are chosen not because they are theoretically optimal, but because they are fast and robust. From this perspective, quantum advantage is unlikely to emerge from direct, one-to-one comparisons with established classical approaches.

If we remain within the realm of electronic-structure calculations, another important distinction emerges: the difference between weakly and strongly correlated systems [18]. In weakly correlated regimes, where mean-field or perturbative methods perform well, quantum computing is unlikely to offer a meaningful advantage. By contrast, strongly correlated systems represent a qualitatively different opportunity. These remain comparatively underexplored in industrial settings. This is not because they lack relevance, but because tools capable of treating strong correlation with both sufficient accuracy and computational efficiency have historically been unavailable. Here quantum computing has the potential to contribute most meaningfully: not by marginally improving well-served problem classes, but by enabling access to regimes that have been systematically excluded by classical tractability constraints.

Examples such as cytochrome P450 and FeMoco have become standard reference systems in discussions of quantum advantage for chemistry. However, strongly correlated electronic structure problems arise in a much broader range of industrially relevant settings. Transition-metal clusters similar to those found in FeMoco have the potential to play a central role in catalytic processes. One prominent example is the search for alternative routes to ammonia synthesis. The Haber-Bosch process currently produces more than 150 million tonnes of ammonia annually and is estimated to consume around 1–2% of global energy production while contributing substantially to global $CO_2$ emissions. Improved catalysts capable of operating under milder conditions could therefore have significant economic and environmental impact.

Strong correlation effects are also central to many classes of functional materials, including high-temperature superconductors, magnetic materials and Mott insulators. The latter are a particularly striking example as they are predicted by simple band theory to behave as conductors but instead exhibit insulating behaviour due to strong electron-electron interactions. Such phenomena remain challenging for many classical electronic-structure methods. Beyond their scientific interest, these materials are relevant to a range of industrial applications including energy technologies, sensing, information processing and electromagnetic-response materials. In particular, the ability to understand and design materials with tailored magnetic and electromagnetic properties could impact areas ranging from advanced electronics to radar-absorbing and stealth materials.

Similarly, metal-organic frameworks (MOFs) used for gas separation, gas storage and carbon-capture applications often contain transition-metal centres whose electronic structure can influence adsorption, transport and catalytic behaviour.

**PDE Algorithms**

Further along the speedup spectrum lie quantum algorithms for solving partial differential equations (PDEs), where high-degree polynomial speedups can be achieved under favourable assumptions [19,20]. In these cases, quantum advantage typically arises from transforming problems into a large linear system or time evolution for which advantageous quantum solvers exist. As a result, the quantum computation is one step removed from the underlying microscopic physics, and the anticipated advantages are polynomial rather than exponential.

The Quantum Linear Systems Algorithm (QLSA) [21,22] is a quantum algorithm for solving systems of linear equations, where the problem is defined by a matrix and an input vector. Rather than computing the full solution vector explicitly, QLSA prepares a quantum state whose amplitudes encode the solution. This representation allows the efficient estimation of global properties of the solution, such as expectation values, without reconstructing the entire vector. For matrices that are sparse, well-conditioned and efficiently implementable on a quantum computer, QLSA can offer a significant computational advantage over classical approaches for certain large-scale problems.

Partial differential equations underpin a wide range of modelling tasks in physics and engineering, yet large-scale, high-fidelity simulations remain among the most demanding workloads in classical high-performance computing. The central bottleneck is the trade-off between accuracy and computational cost: increasing fidelity requires finer spatial and temporal resolution, leading to very high-dimensional systems (PDEs are typically three dimensional in space and one dimensional in time) whose solution cost grows rapidly with system size and physical complexity.

Computational Fluid Dynamics (CFD) is a prominent example for employing PDE solvers in practice. For those, the trade-off has led to a hierarchy of classical approaches. Here, direct Numerical Simulation (DNS), which resolves all relevant length and time scales, provides the most faithful description of fluid dynamics [23]. However, DNS is computationally prohibitive for most practical systems. As a result, engineers rely on reduced models such as Reynolds-averaged Navier–Stokes (RANS) and Large Eddy Simulation (LES), which achieve computational feasibility on classical computers by modelling or averaging over parts of the turbulent flow rather than resolving all scales explicitly. Many problems of practical interest therefore sit in a regime where DNS-level fidelity would be desirable, but only RANS or LES are computationally feasible.

From a quantum computing perspective, potential advantage does not arise from solving nonlinear PDEs directly, but rather is something that can be assessed upon reformulating the problem as a high dimensional linear problem via an embedding method. In several quantum approaches, nonlinear dynamics (such as those arising in Navier–Stokes or lattice Boltzmann formulations) are embedded into large linear systems whose solution dominates the computational cost. Quantum algorithms are naturally suited to this regime, as they can efficiently represent and manipulate very high-dimensional linear spaces, provided that inputs can be prepared and relevant observables extracted efficiently [24]. In this sense, quantum computing may enable DNS-like resolution in regimes where classical methods are currently forced to rely on approximations.

The plausibility of such advantage depends strongly on the physical regime. Problems of low complexity are already classically tractable and offer little incentive for quantum acceleration. Turbulent flows as an example are characterised by high Reynolds numbers and introduce a vast range of interacting scales, creating a more favourable setting for quantum approaches. By contrast, high Mach-number flows involve strong compressibility, shock formation and severe nonlinearity, which reduce the efficiency of linear embeddings and make such problems less natural early targets for quantum computing. For such problems, a more credible progression could therefore start with turbulent but low Mach-number regimes, before assessing the feasibility of fully compressible or hypersonic flows at later stages.

Another challenge is that any quantum approach would have to compete with highly mature CFD software that has been refined over decades by dedicated specialists. Even if quantum methods enabled simulations closer to direct numerical simulation (DNS) fidelity, the relevant question from an industrial perspective is whether the additional accuracy would justify the computational resources, implementation effort and cost required to achieve it.

Across these cases, careful problem scoping is essential. In practice, industrial simulations rarely aim to resolve an entire system at uniform fidelity. Instead, computational effort is focused on specific regions or phenomena that matter most for the question at hand. In the quantum setting, this is not just a practical choice but a necessary one: it determines how large and complex the underlying linear problem becomes, and therefore whether a quantum approach can offer an advantage. As in quantum chemistry, the relevant benchmark is not an abstract asymptotic speedup, but whether quantum methods can realistically improve the trade-off between accuracy and computational cost for problems of practical interest.

**Optimisation Algorithms**

In the case of optimisation problems, quantum algorithms are in general not expected to yield dramatic speedups, but rather modest, low-degree polynomial improvements relative to classical approaches [25]. Broad classes of combinatorial optimisation problems for which exponential quantum speedups are known remain elusive. Instead, many proposed quantum approaches, such as the Quantum Approximate Optimisation Algorithm (QAOA) [[26]], are heuristic in nature and compete directly with an extensive ecosystem of highly developed classical heuristic and exact optimisation methods.

This distinction is well illustrated by the travelling salesperson problem (TSP) [[27]], often cited as a canonical example of computational complexity. While the problem exhibits exponential scaling under exhaustive search like with the Bellman-Held-Karp algorithm ($O(n^2 2^n)$) [[28],[29]], practical instances are rarely approached in this way. Instead, state-of-the-art classical solvers combine branch-and-cut methods, cutting-plane techniques, local search heuristics such as the Lin-Kernighan heuristic [[30]] and sophisticated problem-specific preprocessing to solve instances involving tens of thousands of cities. The relevant benchmark for quantum optimisation is therefore not brute-force enumeration but these mature classical methods that have been refined over several decades.

More generally, for many industrial optimisation problems classical heuristics provide solutions of sufficient quality for practical decision-making. Consequently, demonstrating industrial value requires quantum methods not merely to improve asymptotic complexity, but to outperform highly optimised classical workflows under realistic resource, cost and time constraints.

Combinatorial optimisation algorithms with exponential speedup are to date unknown. And yet, optimisation attracts substantial industrial interest. This apparent discrepancy highlights a more general point: the magnitude of an algorithmic speedup does not necessarily correlate with its economic impact.

**Quantum Machine Learning**

Quantum machine learning occupies an even more uncertain position. Early progress in this area relied heavily on heuristic constructions. Towards more robust and scalable quantum advantage, the field recently increasingly focused on foundational questions, adopting a more principal perspective to investigate the expressivity of quantum models [31], their trainability in the presence of barren plateaus [32], and their ability to generalise beyond training data [33,34]. While these advances clarify necessary conditions for advantage, whether they translate into practical, provable quantum benefit remains an open question. It is also not yet clear whether any future advantages would primarily accelerate model training, inference or both. Practical challenges such as quantum data access and loading, the assumption of available quantum RAM, benchmarking against rapidly advancing classical baselines and performance evaluation further complicate the assessment. Claims of substantial improvements in computational efficiency or energy consumption of QML compared to classical AI likewise remain to be demonstrated for industrially relevant applications.

These differences in algorithmic speedup are not accidental. They reflect how closely a problem is tied to genuine quantum many-body behaviour. The strongest and most robust speedups arise where the computational task itself is quantum mechanical, while more indirect advantages appear when quantum computation is applied to effective or approximate models.

Applications offering only modest computational improvements can nevertheless dominate if they align closely with real decision-making needs or enable problems to be addressed that would otherwise be left untreated. At the same time, any assessment must account for overheads. Hardware constraints, error correction, compilation costs and data-loading overheads can erode theoretical advantages [35,36], particularly for algorithms with limited asymptotic speedup. This is one reason why problem classes with stronger theoretical advantages are often viewed as more promising in the long term: they provide a larger margin within which such overheads can be absorbed. This margin may also allow these algorithms to deliver meaningful value earlier, even on imperfect or resource-constrained hardware.

These considerations also help clarify the limits of near-term variational approaches such as the Variational Quantum Eigensolver (VQE) [37]. Variational methods have played a notable role in advancing the field: they have driven hardware development and deepened understanding of noise and error mitigation. Research across all algorithmic areas remains valuable, particularly in a field that is still learning how to use quantum hardware effectively. Yet, results obtained on small, idealised instances cannot be straightforwardly extrapolated to industrially relevant regimes. For applications that require predictable accuracy and scalability the community increasingly recognises that fault-tolerant quantum computing and algorithms are essential [38]. Methods have progressed from variational techniques like VQE to subspace-based strategies such as sample-based quantum diagonalisation (SQD), which offers higher noise robustness and accessibility of larger Hamiltonians [39]. However, these advances may suffer from the large number of required determinants and expensive classical post-processing [40] and do not resolve the underlying limitations in scalability [41].

The promise is clearly outlined: the existence of asymptotic speedups is not speculative, but mathematically established through well-defined quantum algorithms. This provides a solid foundation for translation into practical applicability.

The author acknowledges the possibility that (despite the described algorithmic promise) the application gap may reflect fundamental physical and mathematical limitations on the extent of achievable quantum advantage rather than shortcomings in translation, integration or organisational structure. This possibility has been discussed in the scientific discourse [42,43]. It is possible that for many industrially relevant problems that show potential for quantum benefit, classical methods will continue to improve sufficiently rapidly or that the costs associated with quantum computation will outweigh any resulting benefits. Determining where meaningful quantum advantages can exist therefore remains an important scientific question. For the purpose of this perspective, however, it is assumed that at least some industrially relevant opportunities for quantum advantage do exist and that in order to find them we have to actually put substantial effort into looking for them. The focus of the discussion is therefore not on proving their existence, but on examining how such advantages, if present, could be translated into outcomes of industrial value.

### (b) Quantum Computing as Deep Technology

Quantum computing is a paradigmatic example of deep technology [44]: technology rooted in fundamental scientific advances that aims to solve major global challenges. A characteristic feature of these is that the route to solution is not instantly evident and often depends on progress across a chain of interdependent components. In this respect, deep tech differs fundamentally from both conventional industrial development and academic research. Unlike traditional industry, it is not sufficient to productise a technology that already functions reliably. Unlike academia, success cannot be achieved by advancing isolated technical elements without regard to how they interact within a broader system. Progress must occur simultaneously from the bottom up, driven by advances in fundamental physics and from the top down, driven by concrete customer and industry needs. These two strands cannot be developed in isolation but need to constantly feed back into each other and align. Like the strands of DNA these elements are co-dependent and mutually expressive: neglecting one strand undermines the function of the whole. Figure 3 illustrates this "deep-tech DNA", emphasising that impact arises from sustained attention to both sides and to their interaction.

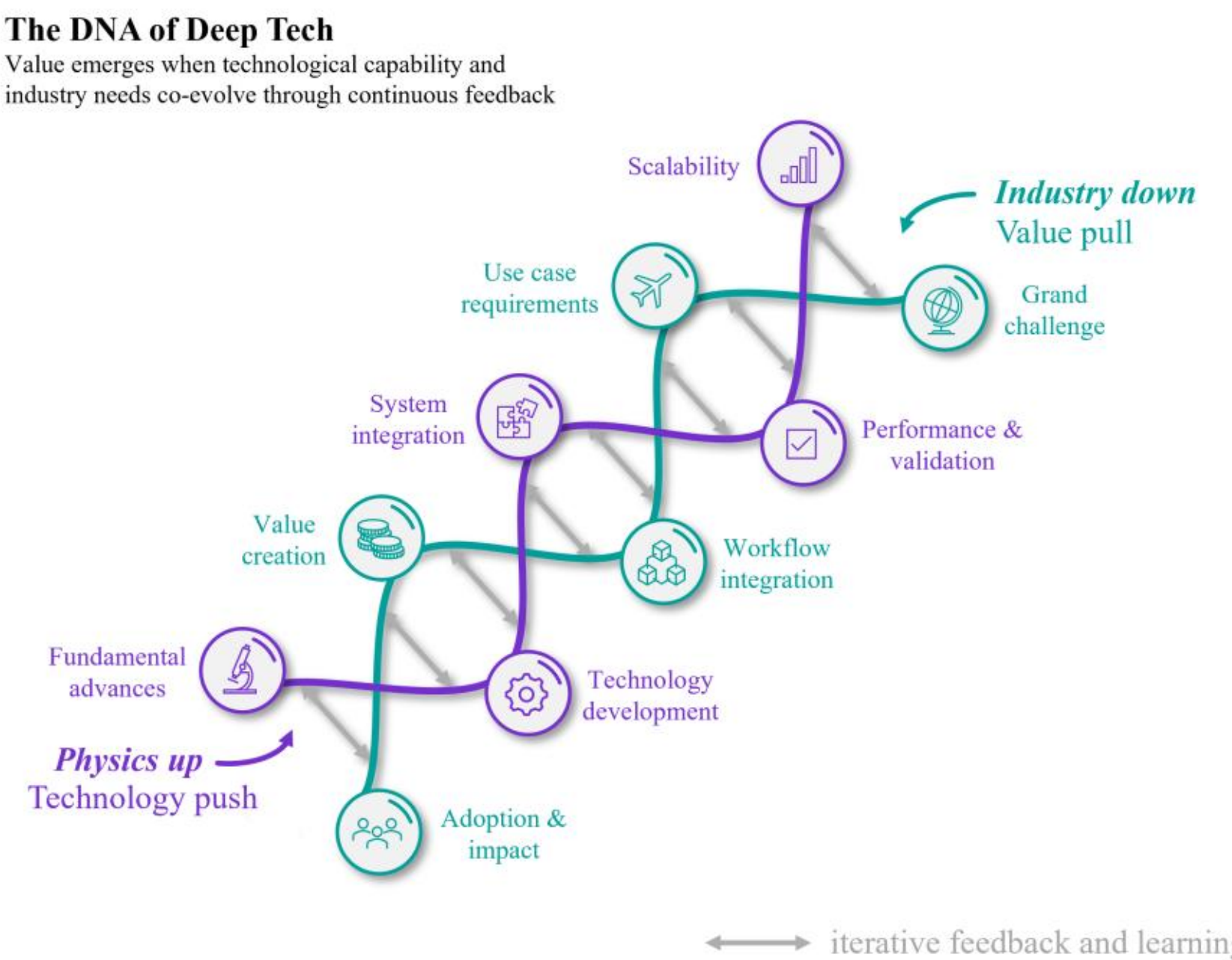

Figure 3: Deep Tech requires development from both sides: first principles and industry need.

Some have therefore advocated an *algorithm-first* approach to application development [10] and this perspective has clear merit: quantum computing cannot address arbitrary problems and it is essential to remain grounded in what quantum algorithms are actually capable of computing. The mathematical foundations of quantum computing place strict constraints on what can and cannot be achieved and any engagement with industry problems must always remain grounded in these realities. However, when pursued in isolation, this approach risks developing in a vacuum, where technically elegant solutions emerge without a credible pathway to industrial impact. Sustained progress requires that algorithmic tractability and application context are considered together, rather than sequentially.

Identifying meaningful quantum computing applications shares features with resolving a *butterfly effect* [45]: the idea that a small perturbation (such as the flap of a butterfly's wings) can, through a complex chain of interactions, ultimately contribute to a large-scale outcome, such as a tornado forming far away. In the context of quantum computing applications, the "butterfly" sits with the algorithm developers, while the "tornado" corresponds to industrial impact.

Between these two extremes lies an extensive and *highly coupled* space. Its structure will depend strongly on the nature of the industrial problem under consideration, and at present it is often not even clear how complex this space truly is. The categories listed below should therefore not be interpreted as a framework or exhaustive decomposition. Rather, they are intended to illustrate the breadth of considerations that may need to be addressed and the extent to which choices made at one level can influence outcomes at another. Ultimately, the key question is whether any advantage obtained from quantum computation survives the cumulative approximations, trade-offs and uncertainties introduced along the pathway to industrial benefit.

Examples of considerations that may arise include:

(a) **Algorithmic level:** Which algorithm should be employed? What level of accuracy or resolution is required? Which basis sets, quantisation schemes, linearisation techniques or factorisation approaches are appropriate? How many quantum computations are required? The optimal choices may differ substantially between systems and applications.

(b) **System definition:** What is the nature of the physical system and which part should be included in the overall calculation? The scale and the areas of interest and thus the system selection and treatment differ significantly between e.g. molecular organic systems in their solvents, large biomolecules like proteins or periodic solid state-systems. For chemical calculations this may involve decisions regarding environmental setup, while in areas such as computational fluid dynamics it may concern the extent of the modelled domain or meshing strategy. Equally important is how the initial system representation is generated and validated.

(c) **Embedding and interfacing:** Where should boundaries be drawn between different levels of theory? How should embedded regions and their interfaces be treated? Where should the transition between quantum and classical methods occur? How are quantities propagated through subsequent processing stages? Each such choice introduces approximations and associated uncertainties.

(d) **Process complexity and resources:** How complex is the overall workflow, and what computational, experimental and human resources are required to implement and maintain it?

(e) **Economic considerations:** What are the costs associated with the various modelling choices and workflow components? What are the costs of required classical resources on top of the quantum computation? Are the expected improvements sufficient to justify these costs in an industrial setting? What trade-offs must be accepted?

(f) **Decision relevance and validation:** How does the quantum-derived information influence the final decision of interest? What benefit remains after all intermediate processing steps? How does this benefit interact with other relevant factors? How can it be validated experimentally or operationally?

The challenge is that this intermediate space is both vast and difficult to survey from any single vantage point. The individuals developing quantum algorithms are necessarily focused on deep technical questions. Expecting them to also anticipate how a given algorithmic capability might propagate through layers of modelling, engineering, certification and market constraints to eventually influence industrial decisions is neither realistic nor fair. The complexity is simply too large and the skill sets required are fundamentally different. Bridging this gap therefore cannot be the responsibility of any single group. It requires deliberate structures, interfaces and collaboration mechanisms that make this translation a shared and continuous process rather than an afterthought.

This is why vigilance, judgement and creativity are all required. We as the quantum computing community must be selective and thoughtful in identifying problem domains where a credible route from quantum capability to benefit can exist, while remaining open to the possibility that the form of that benefit may not be obvious in advance. At the same time, not all domains are equally promising. If one aims to create benefit from QPE calculations of chemical systems, areas such as materials and molecular design offer a more natural alignment than domains like supply chains or network flow.

This also connects to established perspectives in innovation studies. Stokes "Pasteur's quadrant" characterised many transformative technologies as forms of use-inspired basic research, in which fundamental scientific questions and practical needs evolve together rather than sequentially [[46]]. Quantum computing applications exhibit many characteristics of this model. Advances in algorithms are inseparable from questions concerning how resulting capabilities may ultimately generate value. Similarly, the application gap discussed in this perspective shares features with the "valley of death" frequently observed in deep-technology innovation, where scientific advances fail to translate into practical deployment [[47]].

### (c) The Application Disconnect

The promise of quantum algorithmic speedup entered the public discourse in the 2010s [48,49], however, the distinction between specific speedups for narrowly defined problems and general-purpose computational advantage became instantly blurred. The existence of strong asymptotic improvements in a small number of settings was often interpreted as evidence that quantum computers would be able to solve all today's most demanding computational problems, but faster. But even where the narrative did not lose track completely with its mathematical foundation, it was often accompanied by references to broad application areas such as pharmaceuticals, finance or logistics. In many instances this was extrapolated to effortlessly solving some of the civilisation's most pressing problems like climate change or world hunger [50,51].

While these references were not misguided, they typically were vague and largely remain so until today as far as public media is concerned [52]. They rarely gave insight into how problems within these domains could be addressed or how quantum computations could integrate into existing workflows. In the absence of such detail, the dominant message was not about carefully scoped opportunities, but about general-purpose acceleration: that quantum computers would broadly improve established computational practices across entire application areas.

Crucially, this expectation did not depend on the literal truth of any single claim. Even when the most extravagant statements were discounted, the broader narrative still implied universal applicability and fundamentally transformative impact.

As the field matured and the technical community developed a more nuanced understanding of where quantum advantage might realistically arise, the beliefs established by this early narrative nevertheless proved difficult to dislodge. Expectations of broad, automatic applicability continued to influence how quantum computing was discussed, funded and evaluated. As a result, particularly in the early stages of the quantum computing ecosystem the need for deliberate application development was often underappreciated, making sustained investment in this area difficult to secure and leading many organisations to allocate few, if any, dedicated internal resources to application development.

Within the technical community, this public framing was not necessarily taken at face value. Researchers and engineers were generally well aware that quantum speedup had been demonstrated only for specific, narrowly defined problem classes, and that substantial limitations remained. Here the disconnect did not arise primarily because technical experts believed that quantum computers would automatically solve all computational problems. Rather, it emerged from a more structural reality: building quantum hardware and identifying meaningful applications require fundamentally different types of expertise. Teams focused on hardware development are, by necessity, deeply engaged with engineering challenges such as coherence, control, fabrication or scaling.

Hardware development and algorithm design are ideally strongly interfacing, but in their core already those adjacent areas typically proceed as distinct activities, with application considerations forming yet another layer with yet another set of skills and expertise required.

This structural separation was further amplified by the origins of the quantum computing ecosystem itself. Much of the foundational work in quantum hardware and algorithms emerged from academic research groups and many quantum companies were founded directly out of university efforts. This was both natural and necessary. Quantum computing is exceptionally challenging and progress depends on deep theoretical understanding and highly specialised technical skills. In the early stages of the field, such expertise was found almost exclusively in academia, often among young researchers whose training and incentives were shaped by academic modes of working rather than industrial processes.

Academic research typically advances by isolating well-defined problems and pushing the boundaries of what is technically possible within those constraints. This approach is indispensable for scientific progress, but it does not naturally prioritise questions of end-to-end integration. It is often driven by whether a method is more accurate, more efficient or more elegant and as a consequence improvements are often evaluated locally without necessarily considering how those improvements interact with other components of a larger workflow or whether they introduce approximations that may negate gains elsewhere in the pipeline. This is not a shortcoming of the scientists involved, but a reflection of the environment in which they were trained.

As a result, no single group gained full visibility across the entire path from quantum capability to practical use. This separation of expertise did not merely slow progress, it also obscured the true extent of the challenge. Without a holistic view it was difficult to appreciate how much deliberate work would be required to translate quantum computing into meaningful applications.

Indeed, many quantum hardware companies did recognise the need to develop algorithms alongside their devices and invested accordingly by building dedicated algorithm teams. In parallel, a growing ecosystem of quantum software companies emerged. These efforts were essential but would only narrow and not close the application gap.

While discovery of new algorithms is always desired, in practice the main effort of such teams is to make existing algorithms (and thus quantum computations overall) more efficient, scalable and robust. However, these algorithms produce raw computational outputs such as energies, expectation values or probability distributions that are meaningful within a computational framework but remain far removed from what defines real-world applications.

The application gap became more apparent when quantum hardware and algorithm developers and industrial partners attempted to engage directly on application ideas. In these interactions, the term “meaningful” itself was understood in fundamentally different ways.

From the perspective of quantum hardware teams, meaning was shaped by what could be executed on available devices. With only a handful of qubits at their disposal, there was frequently a belief that highly reduced problems might already capture aspects of pharmaceutical or chemical relevance and that running them successfully could begin to deliver tangible value for industrial use cases.

On the other side were industrial experts like computational chemists or pharmaceutical scientists that approached the conversation with a very different set of expectations. They reasonably interpreted the promise narrated by public media as the ability to simulate at least a small-sized protein, accurately and at an electronic level, at short timescales.

Two important insights emerged from these interactions. First, directly mapping molecules of realistic size onto quantum hardware through brute-force approaches was clearly infeasible even in the longer term. Second, even when such calculations could be performed at small scale, isolated quantities such as single-point energies were rarely of immediate use for design-driven problems such as those encountered in drug discovery. Additionally, early concrete resource estimates for electronic energies revealed requirements far beyond any near- or mid-term horizon [53,54]. For industrial partners, such estimates were not an invitation to longer-term collaboration, but a signal of prohibitive cost and risk, leading many to disengage before alternative formulations could be explored.

Significant progress has been made in recent years toward addressing the brute-force challenge by developing methods that identify and isolate those parts of a system where quantum treatment is most likely to add value. This shift was catalysed significantly by the increasing involvement of computational and quantum chemists in quantum computing research, a space that had previously been dominated by quantum physicists. With this broader expertise came established strategies from classical electronic-structure theory for managing complexity in large systems.

In "classical" quantum chemistry, techniques such as active space selection [55], embedding [56] and multiscale modelling [57] have long been used to partition complex chemical systems into regions requiring highly accurate but computationally expensive treatments, and regions where more approximate methods are sufficient. Adapting these ideas to quantum computing has proven both natural and fruitful. However, such partitioning is far from trivial. Interfaces between differently treated regions must be chosen carefully and corrected for energetic and electronic exchange. These challenges are particularly pronounced in dynamic settings, where the nature of the coupling between regions can evolve over time and must be updated consistently.

Well-established classical approaches such as QM/MM methods [58] widely used to model drugs bound in protein environments, have provided important inspiration. Many of these ideas have been carefully adapted to the constraints and opportunities of quantum computing [59,60]. This line of research is widely regarded as essential for the success of quantum computing in chemistry and materials science.

At the same time, these emerging methods introduce new sources of approximation that must be handled with care. Partitioning a system and introducing interfaces inevitably leads to errors, and without rigorous assessment and validation, such errors risk negating the accuracy gains achieved by the quantum computation itself.

The lesson is therefore twofold. On the one hand, problem size reduction represents one of the most promising routes toward actionable chemistry applications. On the other hand, these techniques are not merely technical conveniences, they are design choices that shape how quantum information enters a broader modelling pipeline and thus how applications will be shaped.

This leads to the challenge that the quantum computing community has not yet fully cracked: How to translate to quantities that are of industrial relevance. This requires a look towards what industry requires and what their decision factors are.

### (d) Towards Industrial Benefit

When designing a material or molecule in the industrial setting, value emerges from the ability to satisfy a tightly coupled set of requirements under practical, economic and regulatory constraints.

In materials-intensive sectors, innovation is typically motivated by improvements in capability, reductions in cost or the enablement of functions that were previously unattainable. These drivers translate into concrete and often competing requirements. For example, materials used in spacecraft [61] must combine factors like low mass to minimise launch costs with sufficient stiffness and mechanical integrity to survive launch and operation. They must perform reliably across extreme temperature ranges, exhibit specific thermo-optical properties such as emissivity and absorptivity for thermal control, resist radiation damage in space, minimise outgassing under vacuum conditions, provide appropriate electrical grounding behaviour and withstand corrosion during ground handling and storage. Each of these criteria is critical and improving one can easily come at the expense of others.

Even more complexity arises in advanced materials like composites and layered materials, where interfaces and processing history are as important as bulk composition [62]. Changes introduced to enhance a single property such as stiffness, radiation resistance or thermal stability can lead to unintended consequences including increased brittleness, delamination or compromised interfacial integrity. For industry, these failure modes matter more than idealised material properties, because they determine whether a component can be certified and deployed at all. This also means that for an idealised material property of value for the industrial domain, its impact is not inevitably transformative and translatable to financial benefit.

Industrial design of such materials is predominantly experimental. Where computation plays a role, in many cases, simulations focus on emergent or macroscopic behaviour such as mechanical response, thermal performance or impact resistance rather than on microscopic quantities alone. The question of how quantum computation might contribute is both promising and challenging.

A concrete example can be found in the evolution of stealth aircraft materials and design (Figure 4). Early platforms such as the Lockheed F-117 relied on faceted geometries, not as a stylistic choice but as a direct consequence of computational limitations in the 1970s: available methods could only simulate radar scattering on simplified, planar surfaces, while smooth shapes were computationally intractable. In parallel, radar-absorbing coatings were based on iron-loaded materials, which were effective but heavy, adding on the order of a ton of weight and significantly limiting manoeuvrability and operational flexibility [63,64].

As classical computational capabilities advanced, both aspects evolved. Improved simulation methods enabled the design of smooth, aerodynamically favourable shapes, as seen in later aircraft such as the F-35. At the same time, materials science progressed toward lighter composite coatings with more finely tuned electromagnetic properties, reducing weight while improving durability and performance [65]. This co-evolution of computation, geometry and materials illustrates how advances in modelling capability can directly shape engineering outcomes.

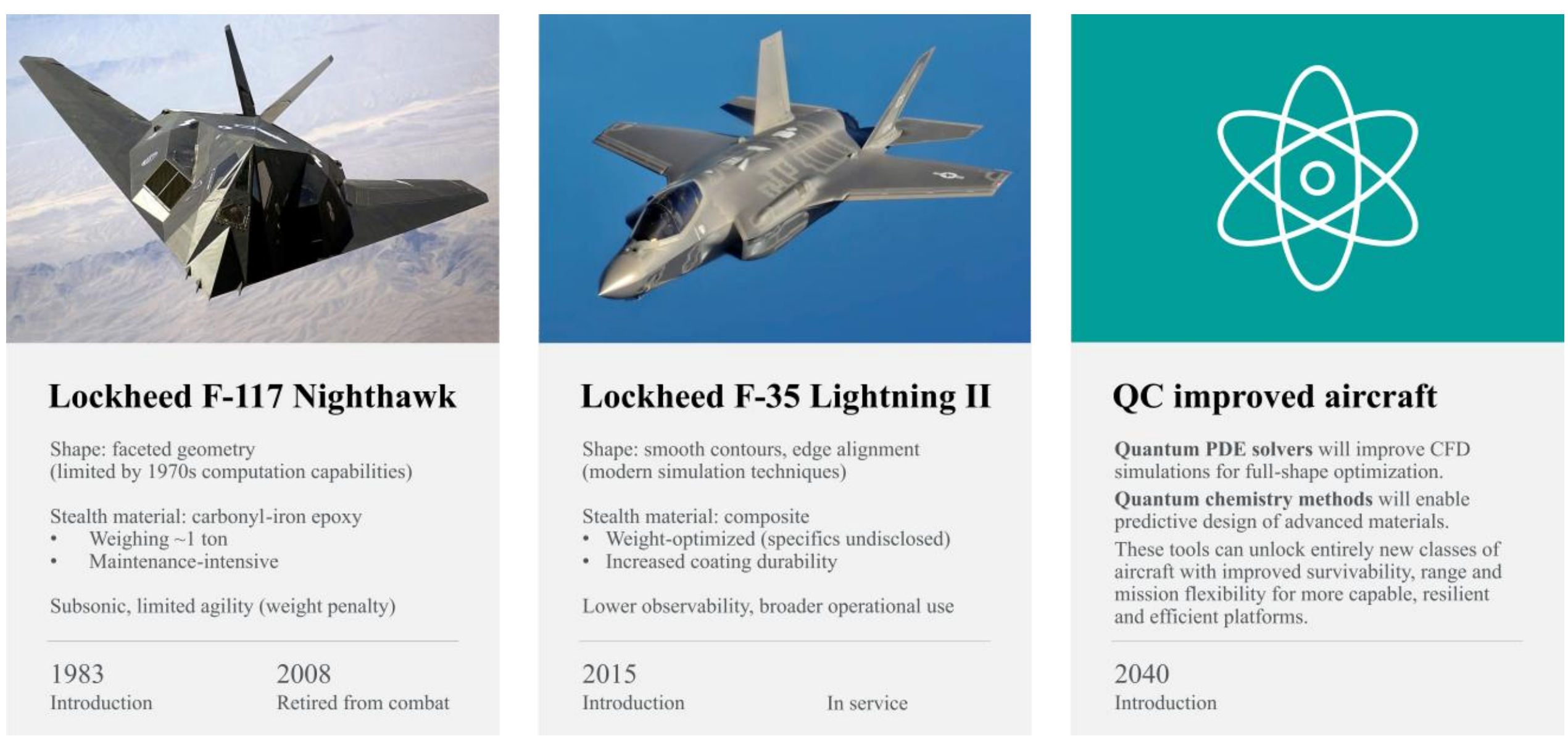

Figure 4: Co-evolution of computation, geometry and materials in stealth design.

Looking forward, quantum computing offers the possibility to extend this trajectory further, particularly in the design of advanced radar-absorbing materials. At the microscopic level, electromagnetic dissipation arises from dielectric, magnetic and ohmic loss mechanisms, which depend on electronic polarisation, charge transfer and spin-dependent interactions [66]. In particular, magnetic losses are governed by quantities such as magnetic susceptibility and spin states, which influence magnetic anisotropy and relaxation behaviour. These are precisely the regimes where strong electronic correlation and spin physics become important, and where classical methods often struggle to provide accurate and predictive descriptions.

From a quantum computing perspective, we need to identify which quantities quantum algorithms could compute reliably, such as spin-orbit coupling and response properties, and understand how these feed into material-level observables. From there, the next the challenge lies in connecting these microscopic quantities to macroscopic performance metrics such as broadband absorption, weight or durability. This requires multiple layers of modelling, from electronic structure to composite behaviour and full-system integration, many of which remain classical and involve their own approximations. An important question is therefore not only how to compute more accurate physical quantities, but how improvements at the quantum level propagate through this chain to influence design decisions. Understanding and engineering this propagation is a non-trivial task, and it is here that much of the remaining work toward industrial benefit lies.

There is a bridge to be built between microscopic quantum calculations and macroscopic industrial concerns. Connecting to the above-mentioned examples: How could quantum chemistry methods be applied to a carefully chosen active space that may provide high-fidelity insight into local adsorption energies and interfacial bonding that aids the understanding whether layers bond sufficiently to each other in order to minimise the risk of material failure under stress. How could partial differential equations help describe large-scale mechanical or thermal behaviour of layered materials of sufficient complexity that both surpasses classical computation limits and is adjusted to quantum capabilities. Building this bridge is not a straightforward translation or substitution of classical elements. Realising it demands deliberate design of multi-scale workflows that combine quantum and classical methods in a coherent way that ensures contribution of targeted, high-value information at points in the pipeline where it can meaningfully inform critical design decisions.

The design of such workflows cannot proceed solely from advances in algorithms. Rather, it requires continuous co-development and iteration across the entire translation layer, including the interfaces between computational outputs, intermediate models and decision-making processes. Equally important is input from the side of the industrial application, where the knowledge resides regarding which quantities are truly decision-relevant, how they influence outcomes, and what forms of improvement would translate into tangible benefit.

**(e) Connecting to the classical Groundwork**

We do not have to look far to learn from an example from classical computing where a computational workflow has been established to inform design. Computer-aided drug design (CADD) [67] in its broader sense is a compilation of computational tools to model, predict and optimise potential drug candidates before they are synthesised and tested experimentally. In the pharmaceutical industry, CADD serves as a critical early-stage filter in the drug discovery pipeline, helping to prioritise the most promising compounds. No single computational result determines success. Instead, decisions emerge from a sequence of interconnected steps, each designed to answer a specific question and to progressively reduce a library of potential compounds down to a few promising drug candidates.

The choice of drug discovery as an example is not arbitrary. At its core, molecular design is a quantum mechanical problem, making it one of the most natural application areas proposed for quantum computing. At the same time, CADD has evolved into a mature computational framework that integrates quantum chemical calculations with a wide range of classical methods. Furthermore, the economic stakes associated with drug development are sufficiently large that even modest improvements in predictive capability may translate into substantial industrial value.

The employed computational techniques rely on different principles: some like docking and FEP (Free Energy Perturbation) [68] are based on evaluating interactions between molecules and protein binding pockets, some like QSAR (Quantitative Structure-Activity Relationship) [69] on statistical models and feature comparison and some like the determination of properties like solubility or toxicity on empirical or data-driven models. These are mainly automated methods that can funnel libraries with millions of compounds. Methods that require intensive manual handling and more computational resources like QM/MM simulations, calculation of NMR spectra or fully resolved reaction mechanisms are employed for bespoke cases and problems and aim rather at gaining insight into processes than acting as routine parts of the pipeline.

From this example, three potential pathways for impactful quantum computing contributions emerge, each representing a different degree of integration and ambition:

**I. Substituting Individual Pipeline Steps:** *One potential approach is to thoughtfully substitute certain CADD pipeline components with quantum calculations capable of computing the same quantities more accurately or efficiently.*

This is an appealing route in principle, but it presents significant conceptual and practical challenges. Many key CADD metrics such as toxicity, solubility or even docking scores are not derived from ab initio quantum chemical methods. In reality, a property like toxicity emerges from complex biological pathways and cellular-level effects, which would require quantum calculations of massive molecular assemblies well beyond quantum capabilities. In cases like docking quantum methods might seem more relevant as describing interactions at an electronic level promises improvement. In practice, the sheer scale of screening millions of molecules makes direct quantum computing evaluation impractical for the time being. CADD was built around classical tools not because quantum methods were overlooked, but because the problem structure and throughput needs align well with fast, approximate, classical models. Therefore, while this pathway could be pursued for specific high-value components, it requires careful identification of well-defined, physically grounded quantities that can be computed better and not just differently by quantum methods.

Alternatively, one could explore whether complex properties like toxicity could be determined through fundamentally different computational pathways with a quantum computation at heart. This would require rethinking how such properties are inferred and whether there exist alternative accessible observables that correlate with biological outcomes. One route could be a thoughtful combination with machine learning approaches.

**II. Substituting DFT:** *Where required accuracy exceeds the capabilities of current classical DFT or wavefunction methods in QM/MM simulations or the study of reaction mechanisms, especially for strongly correlated systems, quantum computing could deliver valuable insights.*

These methods are used sparingly in the pharmaceutical industry, typically for specialised problems where classical models fail or where deep mechanistic understanding is needed. Examples are systems like cytochrome P450 metabolism [70], FeMoco [53] or problem instances like proton-coupled electron transfers [71]. However, these problems sit at the frontier between pharmaceutical industrial and academic research as setting up these simulations is labour-intensive and bespoke. For this pathway to demonstrate repeatable financial impact and measurable value, it will require substantial progress in automation, workflow streamlining and interpretation of data.

**III. Entirely New Workflows for Intractable Classes:** *The most ambitious pathway is to break away from existing CADD pipelines and use quantum computing to enable entirely new classes of drug discovery workflows, particularly for systems that are currently intractable due to their complex electronic structure.*

One example could be the design of photodynamic drugs [72] or excited-state-active compounds, where conventional CADD tools fail due to the lack of electronic structure models that can handle strongly correlated excited states, charge transfer or conical intersections. Here, the idea is not to improve an existing pipeline step, but to construct a new one with quantum computing at its core to tackle a qualitatively different set of problems.

Within this context, the challenge for quantum computing for industrial benefit becomes clearer. In drug discovery as an example, it is quantities like binding affinity, selectivity, reaction rates and kinetic stability that matter and these are emergent properties that depend on ensembles, environments and dynamical processes. Accessing them requires more than accurate electronic energies, it requires computational subroutines within an established workflow.

Encouragingly, recent research has begun to explicitly address this problem. A series of works has explored how quantum algorithms might be structured not merely to compute energies via methods such as quantum phase estimation, but to serve as building blocks within higher-level procedures that estimate related quantities like binding affinities [73], reaction rates [74], free energies [75], NMR spectra [76]

Such efforts represent a significant step toward closing the application gap. They demonstrate that it is possible, at least in principle, to design subroutines based on quantum computing output that get closer to the quantities used in industrial decision-making. While not closing the gap to industrial applications, this is a vital step in gradually narrowing it.

# Implications for Application Discovery

If the central challenge is not only the development of quantum algorithms but also the translation from quantum-computed quantities to industrial benefit, then this has consequences for how application discovery is organised and supported. While it is still too early to prescribe a standard methodology, several implications emerge from the discussion presented here.

**(a) Organisational level**

Teams should deliberately span the translation layer between quantum computation and industrial decision-making. As the examples discussed above illustrate, such teams should include quantum scientists (or at least people with a reasonable grasp of quantum algorithms), experts in the respective classical domains (e.g. quantum/computational chemistry developers, developers for classical PDE codes), domain experts from the target industry who ideally have worked on problems tackled computationally in that sector and translators who enable the whole process. The role of the latter is not to replace domain or algorithmic expertise, but to ensure that the different perspectives remain connected, that the translation target is continually refined and that approximations, errors and decision-relevant quantities are assessed in a common language. In particular, quantum/computational chemists or similar experts do not develop the algorithms themselves, but work with algorithm developers to explore how and into what the outputs should be translated, and what approximations and errors are acceptable. Their task is to examine each step in the translation from quantum-computed quantities to decision-relevant outcomes, identify where information is gained or lost and determine which quantities ultimately matter in practice.

**(b) Translation level**

At the level of the translation itself, the shared vision should be one that everyone involved actively works towards: a translation pathway that is defined jointly by industrial needs and algorithmic possibilities, rather than by either side alone. Progress cannot proceed only from a clearly articulated industrial objective backwards, nor only from what quantum algorithm developers know can actually be built at the basis. Instead, the translation pathway must emerge from two-way dialogues between the transition layer and both industrial needs and algorithmic possibilities, with all participants contributing to the definition of the target. Starting from an industrial objective, one can identify the decisions that influence it, the quantities informing those decisions and the computational and experimental processes through which those quantities are classically obtained. Starting from the algorithmic side, one can identify what kinds of outputs, approximations and error structures are realistically available. Everyone involved needs to be heard. The resulting pathway is unlikely to be known in advance and must therefore be discovered iteratively. Understanding which information is truly decision-relevant, where bottlenecks occur and whether quantum computations can contribute meaningfully are themselves research questions that emerge during this process. Consequently, the translation layer should be regarded not as a fixed pipeline to be engineered, but as a domain of investigation in its own right.

**(c) Ecosystem level**

Sustained exploratory research is required. The broader ecosystem must accept that translation is complex, iterative and uncertain and should therefore enable and support this work rather than forcing it into premature demonstration milestones. While demonstrations can play an important role, an exclusive focus on near-term showcases risks favouring problems that are easy to demonstrate rather than those that may ultimately yield the greatest industrial benefit. Funding mechanisms and collaborative programmes should therefore create space for investigations of the translation layer itself, including the identification of decision-relevant quantities, the mapping of pathways from quantum computational outputs to industrial outcomes and the exploration of entirely new application opportunities. This is also important to counter the challenge of customer fatigue and preserve industrial engagement over the extended development times typical of deep technologies. In many cases, the most valuable outcome may not be a demonstration of quantum advantage but a deeper understanding of where meaningful industrial benefit may or may not emerge.

This also requires preserving spaces in which research can proceed without immediate commercial justification. Deep-technology development often depends on exploratory work whose value cannot be predicted in advance and whose timelines are poorly aligned with quarterly business objectives. While industrial engagement is essential for identifying pathways to impact, parts of the ecosystem must remain sufficiently insulated from short-term commercial pressures to investigate foundational questions whose relevance may only become apparent years later.

If one were to define milestones for progress for bridging the application gap, the first would arguably not be a hardware demonstration but could the construction of a full step-by-step pathway linking a quantum-computed quantity to a decision-relevant outcome. Recent work by Holmes and co-workers [73,[77] [78]] represents an important step in this direction by explicitly analysing how quantum-computed quantities may propagate through application workflows. However, identifying a pathway alone is not sufficient. A second milestone could be the quantitative assessment of how uncertainties and errors introduced by approximations and modelling choices propagate throughout the intermediate layer and how they affect the final outcome and whether any theoretical advantage survives this propagation. A third milestone can be the systematic investigation of which modelling choices genuinely matter for decision quality. This includes understanding how choices such as active-space selection, embedding strategies or approximation levels influence not only computed quantities but ultimately the decisions they are intended to support. Only once these questions are understood can quantum resource estimation be meaningfully connected to industrial benefit and used to assess whether a decision-level advantage is achievable and financially affordable.

# Summary and Outlook

Quantum computing sits at the intersection of deep science and applied engineering. This analysis shows that technical promise and even proof of strong asymptotic speedups is only the first ingredient. Industrial impact requires quantum computations to map onto decision-relevant quantities at the right point in a real workflow. That mapping does not yet exist. Bridging it is the central pragmatic challenge for the field.

We have learned important lessons and the next phase must move faster on application readiness. The challenge now is to ensure that application developments mature in step with hardware. Early narratives were a mixture of genuine theoretical breakthroughs and overgeneralised public claims and led many organisations to defer deliberate, application-focused work. Hardware and algorithm groups matured without always maintaining the sustained, integrated engagement with domain experts that is needed to craft deployable workflows.

Closing this gap demands a two-sided strategy with explicit interfaces. On the one hand, the technical community must continue to push hardware and algorithms (the “tech up” axis). On the other hand, we must design and pursue problems that matter to industry (the “industry down” axis): identify decision points, quantify how computational outputs would change actions and make workflows in which quantum computations contribute measurable value. Crucially, progress depends on tight communication between those axes, ensuring that advances in one direction continuously inform and constrain development in the other.

Practically, this requires purposeful organisational and cultural changes within quantum companies and the wider quantum ecosystem. Deep technical talent (quantum physicists, algorithm designers, computational chemists) is the most critical asset for the field. They are the people who understand the fundamental constraints of quantum computation and who will deliver the genuinely hard technical innovations. But to bridge this gap, teams must be built beyond core algorithm development. In particular, dedicated effort is required to develop *subroutines* and to identify how to transfer quantum computing raw output to *quantities of value.* This layer sits between algorithms and applications and demands its own expertise, combining technical depth with problem abstraction and modelling insight.

Above this, there must be interfaces that actively connect technical development with industrial context in both directions. These interfaces require people who understand industrial problems in sufficient depth to identify what truly matters for a decision, and who also possess enough technical insight to translate those needs into well-defined, computable observables. At the same time, they must be able to interpret technical results back into the language of industry, clarifying what a given quantum computation does (and does not) imply for real-world performance help facilitating necessary collaboration. Experience from industry is beneficial here: Not to replace academic thinking, but to help those trained in academic environments recognise which directions matter and how local technical improvements propagate through real workflows.

Crucially, these structures must preserve *space to breathe* for the scientific work itself. The development of quantum algorithms and methods is intrinsically difficult and intellectually demanding, and it must remain anchored in what quantum computers can genuinely compute. At the same time, assessments of quantum potential must remain grounded in comparisons with the latest classical methods, recognising that advances in classical algorithms and hardware continually redefine the practical boundary of quantum advantage. Industry demand should therefore guide and inform technical direction, but not suffocate it. Forcing algorithmic development to conform prematurely to narrowly specified application requirements risks undermining the very advances on which future impact depends. The challenge is not to subordinate foundational science to short-term industrial constraints, but to keep it continuously oriented toward problems where a credible path to benefit exists. Only under these conditions can creative, high-risk technical work flourish and still evolve in a direction that ultimately enables meaningful industrial use.

Importantly, the perspective presented here should not be interpreted as a proposal for a universal application-assessment framework. Rather, it argues that the field remains in a phase of discovery in which the pathways connecting quantum-computed quantities to industrial benefit must first be identified and understood.

Organisations create the most value by anchoring their efforts in a clear pathway to industrial benefit. This means moving beyond a steady cadence of narrowly scoped “quantum advantage” claims and accompanying press releases aimed at near-term visibility, and instead focusing on building capabilities that translate into real-world impact. Doing so requires disciplined problem selection, realistic expectations and sustained investment in meaningful use cases. When effort is directed this way, talent and attention concentrate on advances that genuinely move the field forward, while strengthening credibility with industrial partners.

Beyond individual organisations, the wider quantum ecosystem also has an important role to play. Industry consortia, funding bodies and government initiatives should internalise the complexity quantum computing applications with industrial benefit and their development, recognising that progress depends not only on advances in hardware and algorithms, but also on the sustained effort required to translate quantum-computed quantities into decision-relevant outcomes. They can help sustain interaction between technology developers and industrial domain experts and support the interdisciplinary work required to bridge the application gap. In particular, funding mechanisms should recognise the development of the transition layer as a scientific challenge in its own right.

By closing the application gap, quantum computing can move beyond the “academic window” and from being enormously valuable for science to industry relevance and a broad societal footprint. If we adopt a deliberate, two-sided approach of technical progress guided by rigorous problem selection and supported by hybrid teams that unite scientific depth with industrial judgment, we improve the odds that quantum computing will meaningfully alter how industry designs materials, molecules and complex engineered systems.

But we should always keep in mind: Ultimately, industry does not care how a problem is solved, nor whether a solution has a quantum computation at its heart. It is the quantum computing community that must care about *how* quantum computation fits into this picture and must do the work required to make that fit meaningful.

## Acknowledgments

The perspectives presented in this article have been shaped by many discussions over the last six years. I have learned a great deal from collaborators, colleagues and interlocutors across academia and industry. Over the past months, a series of in-depth discussions have helped to refine many of the arguments presented here. I am grateful for the time, openness and critical engagement of Marta Mauri, Matija Žeško, Thomas Bendokat, Paul Mannix, Matthew R Hennefarth and Sergio Levi. I thank the two reviewers for their constructive and thoughtful feedback. The time and effort they have put in really helped elevating this paper.

# Additional Information

**AI Usage Statement**
Artificial intelligence tools were used as support during the preparation of this manuscript. In particular, a large language model was employed to assist with structuring arguments, refining language and improving clarity and flow of the text. All conceptual content, technical reasoning, interpretations and conclusions are the author's own and the final manuscript was reviewed, edited and validated by the author to ensure accuracy and integrity.

**Competing Interests**
I have no competing interests.

---

---

[62] Parveez, B., Kittur, M. I., Badruddin, I. A., Kamangar, S., Hussien, M., & Umarfarooq, M. A. (2022). Scientific advancements in composite materials for aircraft applications: a review. *Polymers*, *14*(22), 5007. https://doi.org/10.3390/polym14225007

[63] Rich, B. & Janos, L. (1994). *Skunk Works: A personal memoir of my years at Lockheed*. Little, Brown and Company.

[64] Peter Westwick: Lessons from Stealth for Emerging Technologies, *CEST* **2021**.

[65] Wanting Xu,Na Liu,Zhongchen Lu, Recent Progress of Iron-Based Magnetic Absorbers and Its Applications in Elastomers: A Review, *Materials* **2024**, *17*(16), 4058.

[66] Yang Jin, Haojie Yu, Yun Wang, Li Wang, Bohua Nan, Recent Progress in Electromagnetic Wave Absorption Coatings: From Design Principles to Applications, *Coatings* **2024**.

[67] Hassan Baig, M., Ahmad, K., Roy, S., Mohammad Ashraf, J., Adil, M., Haris Siddiqui, M., Kahn, S., Amjad Kamal, M., Provaznik, I. & Choi, I. (2016). Computer aided drug design: success and limitations. *Current pharmaceutical design*, *22*(5), 572-581. https://doi.org/10.2174/1381612822666151125000550

[68] Chipot, C. (2023). Free energy methods for the description of molecular processes. *Annual Review of Biophysics*, *52*(1), 113-138. https://doi.org/10.1146/annurev-biophys-062722-093258

[69] Muratov, E. N., Bajorath, J., Sheridan, R. P., Tetko, I. V., Filimonov, D., Poroikov, V., Opera, T. I., Baskin, I. I., Varnek, A. et al. (2020). QSAR without borders. *Chemical Society Reviews*, *49*(11), 3525-3564. DOI: 10.1039/D0CS00098A

[70] Caesura, A., Cortes, C. L., Pol, W., Sim, S., Steudtner, M., Anselmetti, G. L. R., Degroote, M., Moll, N., Santagati, R. et al. (2025). Faster quantum chemistry simulations on a quantum computer with improved tensor factorization and active volume compilation. *PRX Quantum*, *6*(3), 030337. DOI: https://doi.org/10.1103/yngp-5fpm

[71] Kovyrshin, A., Manawadu, D., Altamura, E., Pennington, G., Jaderberg, B., Brandhofer, S., Nykänen, A., Miller, A., Talarico W., Knecht, S. et al. (2026). Approximate quantum circuit compilation for proton-transfer kinetics on quantum processors. *Physical Chemistry Chemical Physics*. DOI: 10.1039/D5CP04097C

[72] Zhou, Y., Casares, P. A., Dhawan, D., Loaiza, I., Jahangiri, S., Lang, R. A., Arrazola, J. M. & Fomichev, S. (2025). Quantum Algorithms for Photoreactivity in Cancer-Targeted Photosensitizers. *arXiv preprint arXiv:2512.15889*. https://doi.org/10.48550/arXiv.2512.15889

[73] Otten, M., Watts, T. W., Johnson, S. D., Sundareswara, R., Wang, Z., Hardikar, T. S., Heitritter, K., Brown, J., Setia, K. & Holmes, A. (2024). Quantum resources required for binding affinity calculations of amyloid beta. *arXiv preprint arXiv:2406.18744*. https://doi.org/10.48550/arXiv.2406.18744

[74] Nguyen, N., Watts, T. W., Link, B., Williams, K. S., Sanders, Y. R., Elman, S. J., Kieferova, M., Bremner, M. J., Morell, K. J., Elenewski, J. et al. (2024). Quantum computing for corrosion-resistant materials and anti-corrosive coatings design. *arXiv preprint arXiv:2406.18759*. https://doi.org/10.48550/arXiv.2406.18759

[75] Huang, P. W., Boyd, G., Anselmetti, G. L. R., Degroote, M., Moll, N., Santagati, R., Streif, M., Ries, B., Marti-Dafcik, D., Jnane, H. et al. (2025). Fullqubit alchemist: Quantum algorithm for alchemical free energy calculations. *arXiv preprint arXiv:2508.16719*. https://doi.org/10.48550/arXiv.2508.16719

[76] Fratus, K. R., Enenkel, N., Zanker, S., Reiner, J.-M., Marthaler, M., Schmittecker P. (2025) Can a Quantum Computer Simulate Nuclear Magnetic Resonance Spectra Better than a Classical One? *arXiv preprint arXiv: 2508.06448. https://doi.org/10.48550/arXiv.2508.06448*

[77] Nguyen, N., Watts, T.W., Link, B. *et al.* Quantum computing for corrosion simulation: workflow and resource analysis. *npj Quantum Inf* **12**, 27 (2026). https://doi.org/10.1038/s41534-025-01171-1

[78] Watts, T. W., Otten, M., Necaise, J. T., Nguyen, N., Link, B., Williams, K.S., Sanders, Y. R., Elman, S. J., Kieferova, M. et al. (2024) Fullerene-encapsulated Cyclic Ozone for the Next Generation of Nano-sized Propellants via Quantum Computation, *arXiv preprint arXiv: 2408.13244 https://doi.org/10.48550/arXiv.2408.13244*